# Graph Modelling Analysis of Speech-Gesture Interaction for Aphasia Severity Estimation

**Navya Martin Kollapally[1], Christa Akers[1], Renjith Nelson Joseph[2]**
[1] Kean University, [2] Infosys Ltd
nmartink@kean.edu, christa.akers@kean.edu, renjithnj@gmail.com

## Abstract

Aphasia is an acquired language disorder caused by injury to the regions of the brain that are responsible for language. Aphasia may impair the use and comprehension of written and spoken language. The Western Aphasia Battery-Revised (WAB-R) is an assessment tool administered by speech-language pathologist (SLPs) to evaluate the aphasia type and severity. Because the WAB-R measures isolated linguistic skills, there has been growing interest in the assessment of discourse production as a more holistic representation of everyday language abilities. Recent advancement in speech analysis focus on automated estimation of aphasia severity from spontaneous speech, relaying mostly in isolated linguistic or acoustical features. In this work, we propose a graph neural network-based framework for estimating aphasia severity. We represented each participant discourse as a directed multi-modal graph, where nodes represent lexical items and gestures and edges encode word-word, gesture-word and word-gesture transition. GraphSAGE is employed to learn participant-level embeddings, thus integrating information from immediate neighbor and overall graph structure. Our results suggest that aphasia severity is not encoded in isolated lexical distribution rather emerges from structured interactions between speech and gesture. The proposed architecture offers a reliable automated aphasia assessment, with possible uses in bedside screening and telehealth-based monitoring.

## 1 Introduction

Aphasia is commonly caused by cerebrovascular accident (CVA), or stroke, affecting the language centers of the brain, which for most individuals are in the left hemisphere[López-Barroso et al., 2023]. In the United States, approximately 800,000 people per year experience a stroke and approximately one-third of stroke survivors are diagnosed with aphasia [NIH, 2025]. Even though aphasia symptoms may resolve for some individuals, most continue to live with the persistent language impairment[Wilson et al., 2023]. Severity of aphasia ranges from mild deficits to profound impairment of verbal and non-verbal communications skills. There are different subtypes to aphasiac speech based on the fluency factor. Fluent aphasias are associated with damage to the Wernicke's area i.e. posterior area of the left hemisphere, the speech of patients with this aphasia is produced with ease however may be meaningless[Goll et al., 2010]. In contrast, the non-fluent aphasia is associated with damages to the Broca's area i.e. the anterior portion of the left hemisphere. Non-fluent aphasia are characterized by limited and effortful speech with utterances that are reduced in length and with grammatical inaccuracy's[Ash et al., 2010].

Regardless of the classification of aphasia types the patients with aphasia are present with some degree of word retrieval difficulty. When a target word is inaccurately retrieved, the word produced in error is called a paraphasia[Saling, 2007]. Paraphasia is a diagnostic feature of aphasiac speech. There are different type of paraphasia. A phonemic paraphasia also called literal or phonological paraphasia occurs when the word produced in error share a phonemic relationship with target e.g. sink instead of think[Kurowski et al., 2016]. When an unintended real word is substituted for the target, the error is called a sematic paraphasia, since the error word comes from the same sematic category e.g. hammer instead of nail[McKinney-Bock et al., 2019]. In some cases, they may be loosely connected or may not belong to same semantic category e.g. pillow instead of bread. Neologistic paraphasia or neologism, occurs when a non-word which shares no semantic or phenomics relationship [Rohrer et al., 2009]with the target, for example saying 'krumplik' instead of spoon. Though SLPs commonly use formal assessments like the WAB-R when evaluating persons with aphasia, such tools are criticized for measuring discrete linguistic skills, not everyday language.

Generally, speakers with more severe aphasia experience more frequent word finding difficulties. Word finding errors can affect any lexical class (e.g., nouns, verbs, adjectives, adverbs). Speakers with nonfluent aphasia, particularly those with syntactic deficits known as agrammatism, may be more susceptible to verb retrieval difficulties. Speakers with fluent aphasia types may be more prone to noun retrieval difficulties. However, some researchers have suggested that the relationship between aphasia type and lexical class is not consistently observed [Faroqi-Shah et al., 2003, Hepner et al., 2020]. Gestures (e.g., arm, hand, and body movements that express ideas and intentions), are found in everyday communication. Gesture and language are intrinsically related[Dipper et al., 2015]. For speakers with and without aphasia, gesture use can improve word retrieval during spontaneous speech. Additionally, speakers with aphasia may use gestures to compensate for their word communication impairment. Kong et al. [Kong et al., 2015] found that greater aphasia severity was linked to increased gesture use during discourse production. Gesture use, therefore, may be linked to WAB-AQ scores.

Discourse analysis explores how persons with aphasia use language beyond the sentence level to communicate in functional, real-world contexts[Rohrer et al., 2009, Kuvač Kraljević et al., 2023]. Even though, discourse analysis can be providing rich information, the time required to collect, transcribe, and analyze discourse samples can be prohibitive in many clinical settings[Alyahya, 2025].

Accessing aphasiac severity from a spontaneous patient discourse is challenging due to the complex interaction and multi-modal behavior. In this work, we introduce a state-of-the-art (SOTA) graph neural network modelling for aphasia severity assessment. The proposed graph structure captures observable language behaviors such as speech structure and gesture use, and clinical feature like paraphasia pattern across patient discourse. From this representation we model graph, linguistic and error-based paraphasia features and learn discourse embeddings to predict the aligned WAB severity score called as WAB Aphasia Quotient (WAB-AQ).

We are performing two complimentary tasks here, in task one we perform quantitative analysis using correlation analysis, multivariate ridge regression, heteroscedastic- robust (HC3) estimation [Jochmans, 2022] and an ablation study. This multi-step study is to understand the contribution of graph structure, part-of-speech distribution and paraphasia types as stand-alone and combined factors for aligning with the gold-standard WAB-AQ scoring for each patient discourse available from the Apahsia Bank[Forbes et al., 2012]. In the second task, we train a graph-neural network to learn the discourse level embeddings and predict the WAB-AQ score. Finally, we also perform a supplementary qualitative analysis using out-of-fold prediction to understand the best, median and worst representative cases. Results show that graph-based learning provides the greatest alignment to the clinical scoring and adding gestures, POS (part-of-speech) and paraphasia information gives a reliable and extensible framework to predict the WAB scores.

The main contributions of our work are as follow:

1) We introduce a graph-based representation of aphasiac discourse using weighted interaction graph with nodes representing POS tags, gestures and paraphasia events and edges represents the frequency of interaction between edges.

2) We present a node-level feature encoding and discourse level derived embedding to model the multi-modal discourse for aphasiac speech.

3) We present a quantitative analysis using spearman correlation, HC3 regression and ablation study to understand the contribution of graph structure, linguistic and paraphasia contribution to severity prediction.

4) We use the graph embeddings and train a graph neural network to predict the WAB scores.

5) We present a qualitative analysis using out-of-fold prediction and discourse graph visualizations.

6) We introduce and end-end framework multimodal aphasiac analysis from transcript processing to neural modelling using the patient discourse available in Apahsia Bank.

## 2. Background

In aphasia assessment, Artificial Intelligence (AI) has commonly been used for differential diagnosis (e.g., aphasia versus no aphasia) and for determining aphasia type [Azevedo et al., 2024]. For example, Konstantinopoulou et al. [Konstantinopoulou et al., 2019] used a multilayer feed forward neural network to explore 19 attributes in 164 records from AphasiaBank to classify aphasia types. Le and colleagues [Le et al., 2018] used an automatic speech recognition (ASR) system to analyze spontaneous speech samples from AphasiaBank. The authors produce a set of features for each speaker that are characteristic of aphasic speech, compatible with ASR output, and robust to recognition errors. They analyzed the lexical structure based on Information Density (DEN), Dysfluency (DYS), Lexical Diversity and Complexity (LEX), Part-of-Speech Language Model (POS-LM), Pairwise Variability Error (PVE), and utterance, Posterior Gram-Based Dynamic Time Warping (DTW). They determine that acoustic modeling could accurately predict WAB AQ scores without manual transcription.

TalkBank, renowned as the world's largest open-access repository of spoken language, plays a pivotal role in advancing linguistic research and understanding language disorders[MacWhinney, 2025]. AphasiaBank, a key initiative within TalkBank, is a comprehensive multimedia database specifically designed to study and educate about language loss resulting from Aphasia[Forbes et al., 2012]. The database contains 402 hours of interviews with individuals with Aphasia and 220 age-matched control participants. It also includes transcripts of the video linked at the utterance level with the video. Standardized protocols are implemented to ensure consistent data collection across different research studies[Forbes et al., 2012]. They include personal narratives, picture descriptions, story retelling, and procedural discourse. The English section of the AphasiaBank database is divided into four sections: Aphasia, Controls, Non-Protocol, and Script.

transcribed in CHAT format and linked line by line with video.

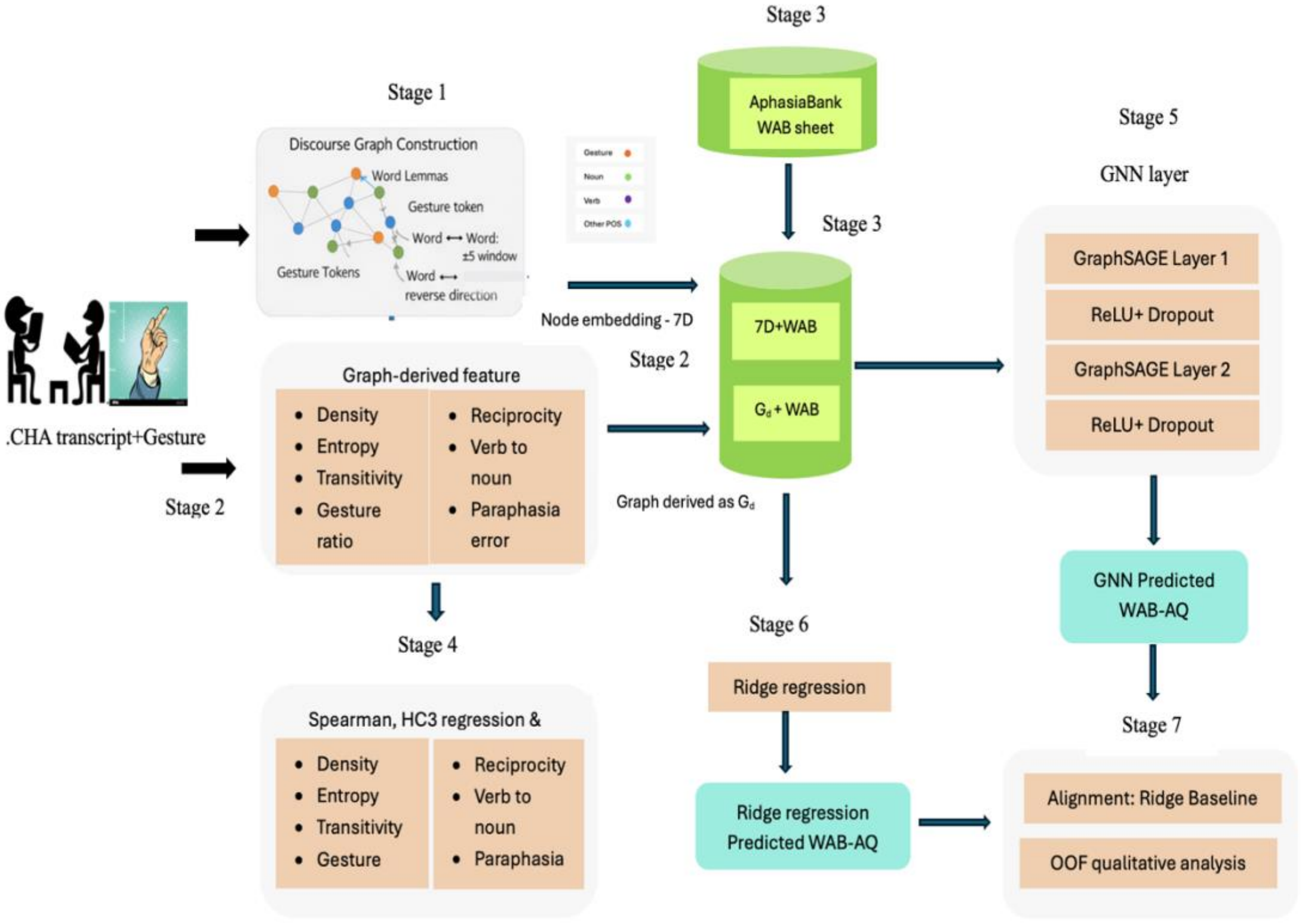

Figure 1: Architecture diagram of aphasia severity prediction and analysis using graph modelling

## 3. Methodology

We have divided the framework to seven stages as follow. Figure 1 gives the architectural flow diagram of the framework with arrows indicating the process and output points.

Stage 1 Data pre-processing and Graph construction

The input data consist of transcripts(.CHA file) from 192 participants with a total of 287 hours long of interaction with SLP in Aphasia Bank under Non-Protocol section for English speakers. These interactions consist of patients talking about personal narratives, picture descriptions, story retelling, and procedural discourse like making a peanut butter and jelly sandwich. We focused on mixed participant pool with patients at different aphasia levels as no aphasia, mild, moderate, severe and very-severe aphasia.

The non-protocol section contains samples of people with aphasia engaged in a wide range of communicative tasks and interactions, valuable for speech-language pathologists (SLPs) because it illustrates diverse patterns of aphasic communication. For example, interviews with stroke survivors in both acute and post-acute stages, describing their experiences of stroke, Aphasia, and adaptation[Forbes et al., 2012]. All materials in the non-protocol database are transcribed in CHAT format and linked line by line with video.

In this work, we focus on the non-protocol section of AphasiaBank and utilizes the WAB scoring sheet provided in AphasiaBank as the gold standard for comparison. The 192 samples out of 512 participant samples were selected as it consists of no null and unknown values for WAB scoring, WAB spontaneous fluency scoring, WAB sequential command scoring and WAB repetitions in the WAB score sheet. This subset of scores were chosen as recommended by SLP to make sure that all regression and cross-validation are done consistently across samples avoiding bias and instability from missing values.

From the transcripts participant conversations were extracted using the tag [*PAR], for every conversation the script extracted [%mor], [%wor], [&=] for morpho-syntactic (parts-of-speech tag), word-level tokens and gestures respectively. Gestures in AphasiaBank are not linked to words. We approximate the relationship between gestures and words, by pairing each gesture with a nearest word token within five tokens on either side of the gesture. This results in a directional relationship depending on whether the gesture or word comes before i.e. whether the gesture precedes or follows speech. This is one of the ways we capture the temporal alignment between multiple modalities in each patient discourse. The paraphasia errors were extracted using following patterns [*s:] for semantic paraphasia, [*p:] for phonemic paraphasia and [*n:] for neologistic paraphasia. Other pre-processing steps included removing error markers like [\+] and [*], timestamps like 123_456, stray digits and punctuations. After the pre-processing steps we used Louvain community detection method to construct a graph per patient discourse. In the graphs as in Figure 2 the orange nodes indicate gestures, green nodes are nouns, purple nodes are verbs, and light blue nodes are other POS category. Each node is labelled, and the frequency indicates the count of how often that node appeared in conversations. Edges indicate relationship between word-word, gesture-word and word-gesture. Each edge weight indicates the number of occurrences of that relationship between the connecting nodes.

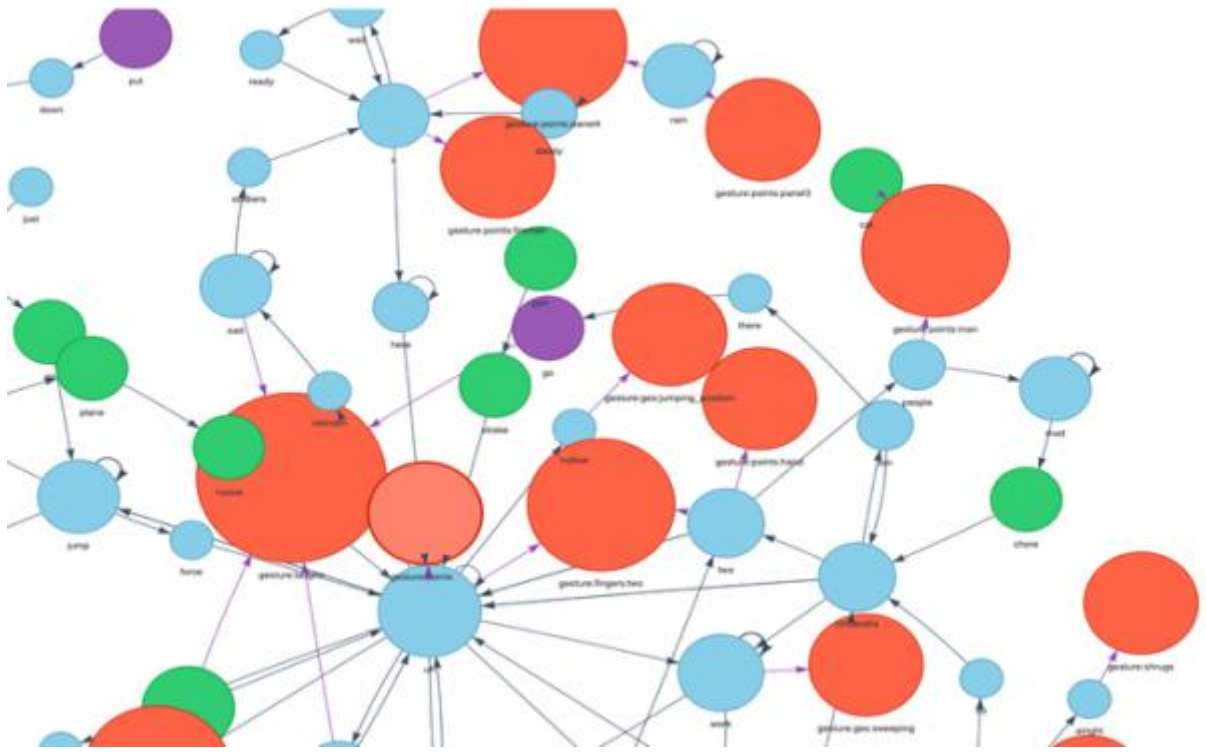

Figure 2: Louvain graph constructed for each patient discourse.

Stage 2 Feature extraction from transcript and graph

After constructing the multi-modal graph for each transcript, we derive two parallel representations. First, is graph derived features as $G_d$, for each graph that was generated per patient. Let graph (G) be defined as (V, E) where |V| indicate the number of nodes as n and |E| indicate the number of edges as m. For each of the graph we calculate the graph features such as density, reciprocity as the tendency for bidirectional links, transitivity as clustering in the undirected version, proportion of gesture node to quantify gesture vocabulary size, degree entropy of the graph to understand the diversity of nodes. We also calculate the edge weights statistic and fractions as defined below and paraphasia error features per graph in terms of semantic, phonological and neologistic error rates.

1. For graph G(V, E) we define |V|= n where n is the number of nodes and |E|= m where m is the number of edges.
2. Density of graph $= \frac{m}{n(n-1)}$
3. Reciprocity $= \frac{|\{(i,j)\in E,\ (j,i)\in E\}|}{m}$
4. Transitivity $= \frac{3*\text{number of closed triangles}}{\text{number of connected triples of nodes}}$
5. Gesture to node ratio $= \frac{\text{number of gesture node}}{\text{number of nodes}}$
6. Degree entropy $= -\sum_{i=1}^{n} \frac{d_G(V_i)}{\sum_{j=1}^{n} d_G(V_j)} \log \frac{d_G(V_i)}{\sum_{j=1}^{n} d_G(V_j)}$

Where graph vertex set of $V = \{V_1, V_2, ..., V_n\}$, and let $d_G(V_i)$ be the degree of $V_i$ of G.

7. Edge weight statistics- Summarized using the average and maximum values of different edge types as W-W, G-W and W-G; W showing word and G for Gesture.
8. POS features- Noun count, Verb count, noun rate, verb rate. Noun to verb ratio.
9. Paraphasia error rate- Overall proportion of words tagged with the type of paraphasia to the total number of word tokens as para_any. para_sem is the proportion of tokens replaced with semantically related word similarly we measure para_phon for phonological paraphasia and para_neo of neologistic paraphasia.

In the second representation for learning each node is encoded using a 7D feature vector and this representation is called node-level feature vector. The node-level feature vector is composed of log-scaled node to reduce the dominance of high-frequency nodes with discourse being in varying length from 12 minutes to 42 minutes, one hot encoding for token-type (Gesture vs Word), one hot encoding for POS (Verb, noun, other POS-category).

Stage 3: Graph and Clinical feature merging

In this stage, clinical assessment variables from the Western Aphasia Battery (WAB) are merged with the node-level and graph-derived features extracted from Stage 2 to form a unified representation for the GNN modeling and evaluation. The WAB Aphasia Quotient (WAB-AQ) is used as the regression target, as it provides a continuous measure of overall aphasia severity. In addition to WAB-AQ, three WAB subtests, Spontaneous Speech Fluency, Sequential Commands, and Responsive Speech, are retained as additional clinical variables because they are most directly related to discourse-level language production and comprehension. The result of this stage is an integrated feature representation per participant, providing appropriate inputs and targets for graph-based regression models and ridge regression modelling.

Stage 4: Quantitative analysis and ablation study

We use the graph-derived feature vector $G_d$ unified with WAB-clinical scoring obtained in Stage 3 to perform the statistical modelling at this stage. We first compute Spearman correlation heatmap as in Figure 3 to identify the correlation between each input feature with WAB-AQ and other WAB scores. Since the relationship between graph-derived features and WAB-scores may or may not be linear we use a Lowess smoothed plot to visualize the potential non-linear relationship between graph-derived feature and WAB scores. As in regression estimates the distribution shows variance around the fitted line for each data point, we perform multivariate regression model using HC3 in which this assumption falls. In HC3 standard errors are used to make the regression robust to the heteroscedasticity in the residuals. At this stage we perform the first ablation study to quantify the impact of gesture and paraphasia rates to the score prediction.

Step 5 Graph Neural Network Modelling

Each discourse graph along with the 7D node-level feature vectors with clinical WAB scores as target column are converted to a PyTorch geometric object and saved as .pt file for training the GNN. Our model consists of two GraphSAGE convolutional layers followed by a ReLU activation and dropout for regularization. Training uses mean squared error loss and Adam optimizer. The performance of the model is evaluated using 5-fold cross- validation. At this sub-stage the goal is to understand whether the graph-neural network can automatically learn the discourse graph representation without adding graph-derived features. Also, at this stage we check the model performance excluding the gesture information to see if adding gestures actually help towards predicting the WAB scores.

Step 6 Ridge Regression Baseline

To compare the GNN model performance we use a classic machine learning model a ridge regression model. The input to the model is the graph-derived feature vector and WAB scores are the target labels ($G_d$+ WAB). The performance is evaluated using a 5-fold cross-validation and reported using same metric as stage 5 i.e. root mean squared error (RMSE) and mean absolute error (MAE).

Step 7 Qualitative analysis of performance

In this final stage we perform a qualitative analysis to interpret the GNN's model performance in WAB score prediction at the discourse level. Using the saved PyTorch graphs we retain the GraphSAGE regressor under repeated 5-fold cross validation and collect the out of fold (OOF) predictions for every participant, ensuring the model did not see the participant data during training. We aggregate the predictions by averaging the OOF estimate per participants. Table 2 shows the best, worst and median error per cases.

## 4. Results

In our analysis semantic and neologistic errors had strong correlation between aphasia severity and type. Figure 3 shows the relationship between paraphasia type and probability of aphasia. Overall paraphasia rate defined as 'para any' and semantic paraphasia indicated as 'para sem' have a clear positive association with aphasia. As the paraphasia rate increases the probability of being in sever aphasia rises with odd ratio of 1.25 and 2.80 per 0.01increase for para any and para sem. In contrast phonological paraphasia as 'para phon' shows little to no positive effect in aphasia type. Across the graphs in Figure 2 for different paraphasia, non-aphasic speakers cluster near zero paraphasia rates.

From the spearmen correlation analysis present in Stage 4 we observe that density is negatively correlated with degree entropy, aligning with the hypothesis that denser graphs tend to have more uneven connectivity. Also, POS features such as noun rate, verb rate were negatively correlated with WAB score as in the heatmap Figure 4. Most of the individual graph features showed a weak correlation with WAB score.

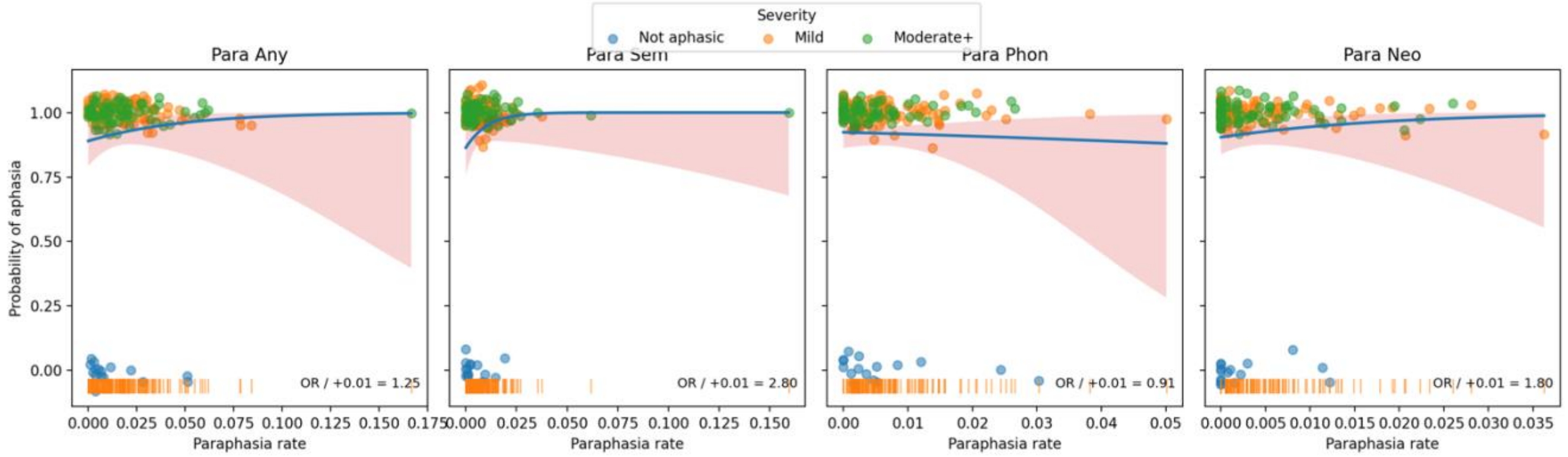

Figure 3: Paraphasia rate for combined, semantic, phonological and neologistic vs aphasia probability.

One important observation from Figure 4 is that gesture ratio has a negative association with WAB score indicating that higher reliance on gesture is linked to severe aphasia. Similarly, paraphasia rate shows a small negative correlation indicating a trend that cannot be generalized, in terms of increased errors indicating a lower WAB score. This analysis also affirms the fact that no single feature can predict aphasia severity and motivating the use of a multivariate regression model combining multi-modal feature input.

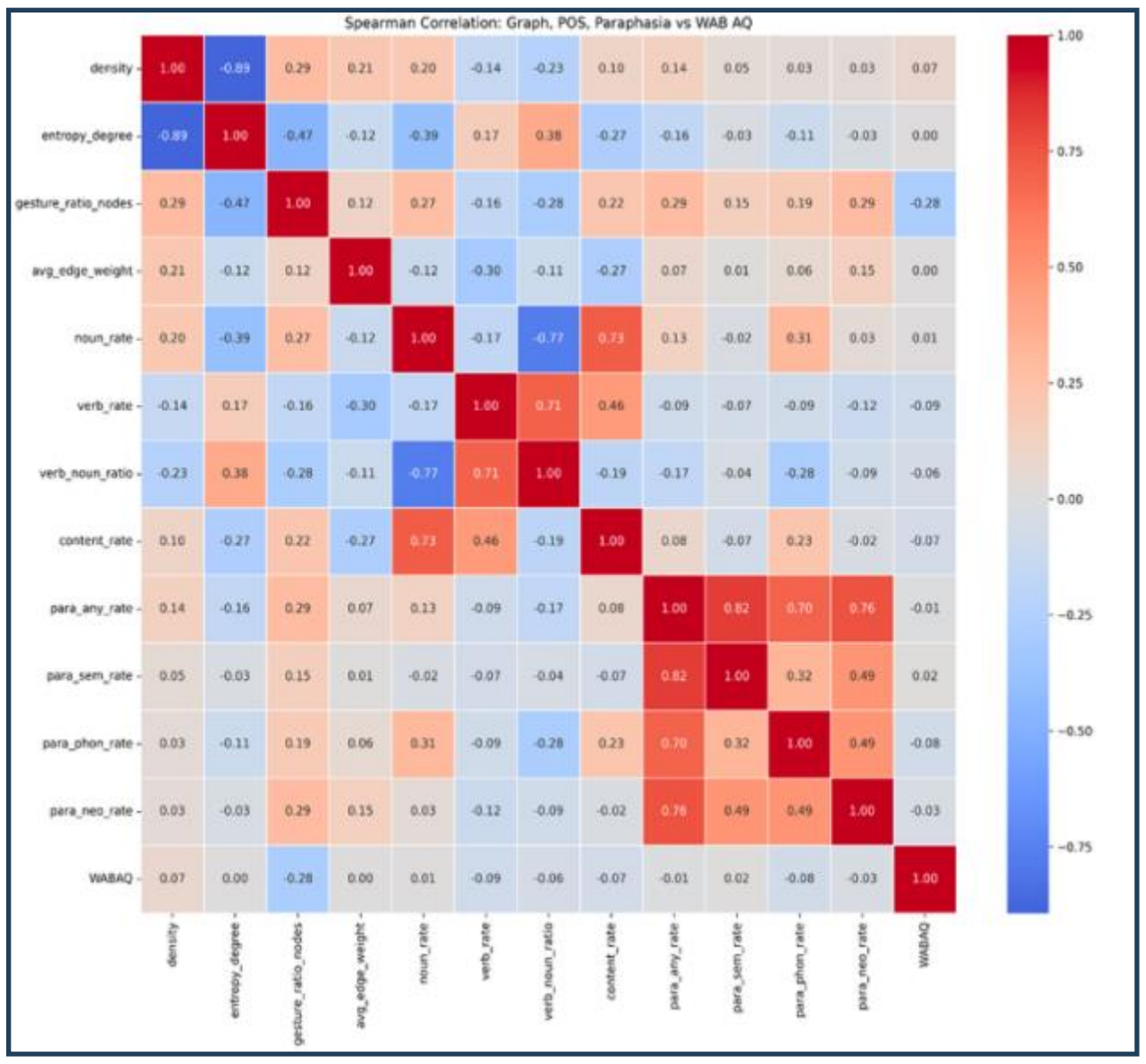

Figure 4: Sperman correlation with graph-derived metric.

From the ablation study we observed that graph only model performs substantially better with $R^2$ value of 0.089. Adding POS features to the model improves the predictions with $R^2$ value of 0.099. Combining graph, pos and paraphasia features we obtain an $R^2$ value of 0.120 and adjusted $R^2$ value of 0.087. POS only and paraphasia only models had an $R^2$ value of 0.002 indicating that relaying just on grammatical or paraphasia error rates is insufficient to predict aphasic severity.

Table 1 shows the output of Stage 5 and 6 and alignment between the GNN and ridge regression. The GNN achieved a RMSE= 9.29±1.47, Pearson r =0.703±0.082 displaying strong alignment between predicted and true aphasia scores. Performance is also reasonable for Spontaneous Speech fluency, Sequential Commands and Repetition. These results show that using the 7D feature vector for each node the GNN can learn meaningful representation from the discourse graph.

Ridge regression model trained on graph-derived, POS and paraphasia features also achieved a substantial correlation with WAB score but it relies on multi-modal calculation involving graph, pos and paraphasia factors. GNN learned directly form graph topology when ridge regression relied on explicit feature engineering. Ablation study while modelling GNN showed that removing gesture related nodes reduced performance achieving RMSE= 18.9±1.47, Pearson r =0.67±0.082. This supports the earlier finding that gesture usage is a reliable indicator of aphasia severity.

From the Figures 5 and 6 it shows the predicted WAB vs true WAB for GNN and ridge regression. The reported Pearson and Spearman for GNN indicates a strong linear agreement between predicted and true score. Larger spread at low WAB shows that the model struggles at very severe aphasia cases, where there are more variables and sparse graph. Points are widely scattered especially at mow and mid WAB for ridge regression. There are extreme outliers in the form of large negative predictions indicating the model is sensitive to unusual feature combination for severe aphasic cases.

| Target | Model | RMSE | MAE | Pearson r | Spearman ρ |
|---|---|---|---|---|---|
| WAB-AQ | Ridge | 16.42±4.76 | 5.78±1.56 | 0.703±0.093 | 0.716±0.068 |
| | GNN | 9.29±1.47 | 11.52±1.38 | 0.703±0.082 | 0.675±0.094 |
| Spontaneous Speech fluency | Ridge | 1.77 ± 0.17 | 1.48 ±0.12 | 0.712±0.081 | 0.692±0.083 |
| | GNN | 1.84±0.21 | 1.53±0.18 | 0.650±0.104 | 0.632±0.105 |
| Sequential Commands | Ridge | 8.321 ±3.48 | 6.11±2.18 | 0.504±0.094 | 0.504±0.100 |
| | GNN | 7.501±1.946 | 6.06± 1.95 | 0.471±0.106 | 0.461±0.088 |
| Repetition | Ridge | 2.83±0.63 | 1.92±0.17 | 0.524±0.142 | 0.535±0.107 |
| | GNN | 2.49±0.24 | 1.84±0.18 | 0.573±0.121 | 0.528±0.084 |

Table 1: Ridge regression vs GNN alignment from Stage 5 and 6

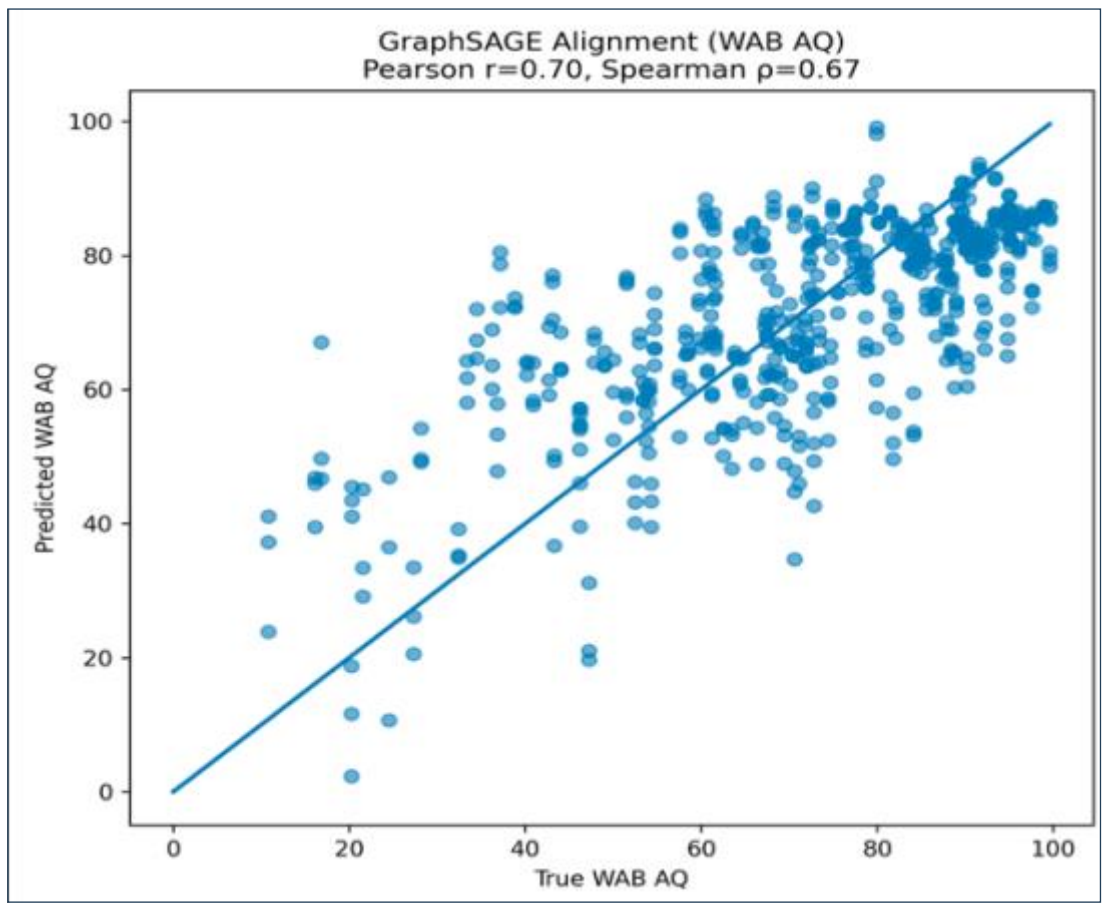

Figure 5: Predicted vs True WAB-AQ for GNN modelling

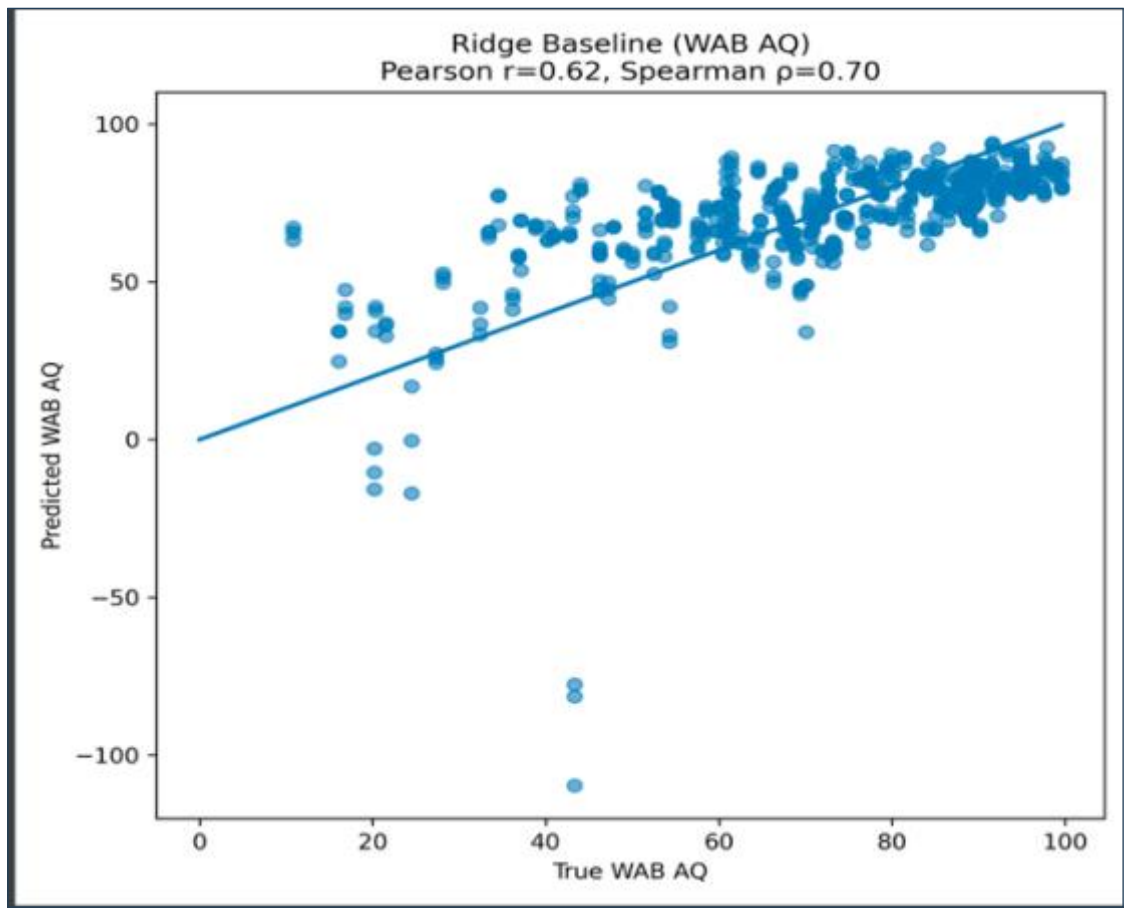

Figure 6: Predicted vs True WAB-AQ for ridge regression modelling.

Table 2 reports the best, median and worst cases based on the OOF prediction modelling in Step 7, it's clear that the worst cases align with the models struggle in predicting very sever aphasia type where the discourse graph is parse. The model achieves near perfect results for moderate aphasia types but fails on small number with large overestimation. The results highlight that graph derived and linguistic feature along with paraphasia rate jointly improves the predictions. However, the performance degradation with very sever aphasia highlights the limitation of graph-based learning in a low-information setting.

| True WAB-AQ | Pred WAB-AQ | ABS error |
|---|---|---|
| 78.70 | 78.35 | 0.35 |
| 59.70 | 60.06 | 0.36 |
| 67.20 | 66.81 | 0.39 |
| 70.60 | 75.18 | 4.58 |
| 68.60 | 73.42 | 4.82 |
| 84.10 | 79.20 | 4.90 |
| 64.50 | 83.00 | 18.50 |
| 99.59 | 81.37 | 18.22 |
| 96.09 | 78.39 | 17.70 |

Table 2: Best, median and worst WAB-Q from Stage 7 OOF prediction

## 5. Conclusion

This study proposed a graph-based modelling for estimating the severity of aphasia integrating speech, gesture and POS features. By correlation analysis, robust regression with HC3 standard error and ablation studies we showed the importance of multi-modal learning using combined feature input. Ablation studies indicated that graph only model captured aphasia severity but adding linguist and paraphasia features improves the performance.

We further demonstrated that GraphSAGE model learned directly form the discourse graph, achieving a performance comparable to the ridge regression model trained on graph-derived features. Out of fold qualitative analysis indicated that the model performance well for mild-moderate aphasia and struggles at very sever aphasia. During very sever aphasias the discourse graph becomes extremely sparse revealing an important limitation of graph-based modelling in low-information setting.

## 6. Future Work

Future work will focus on improving predictions in very sever aphasia setting where discourse graph becomes sparse. We aim to incorporate acoustic features and temporal speech dynamics for very sever aphasia setting. In addition, attention-based graph modelling and multi-modal fusion strategies will; be investigated to improve the performance. We also plan to extend this study in multi-lingual setting.